\documentstyle[12pt]{article}

\def\simlt{\stackrel{<}{{}_\sim}}
\def\simgt{\stackrel{>}{{}_\sim}}

\def\be{\begin{equation}}
\def\ee{\end{equation}}
\def\bear{\be\begin{array}}
\def\eear{\end{array}\ee}
\def\bea{\begin{eqnarray}}
\def\eea{\end{eqnarray}}

\def\c2bb{\cos^2 2 \beta}

\def\s2bb{\sin^2 2 \beta}

\def\simlt{\stackrel{<}{{}_\sim}}
\def\simgt{\stackrel{>}{{}_\sim}}
\def\ov{\overline}

\def\tb{\tan\beta}

\def\matb{(m_A,\tan \beta)}

\def\st{\tilde{t}}

\def\tev{\rm \; TeV}
\def\gev{{\rm \; GeV}}

\begin{document}
\begin{titlepage}
\thispagestyle{empty}

\title{{\bf The Electroweak Phase Transition in Extended Models}
\thanks{Talk given at the
NATO Advanced Reseach Workshop: {\em Electroweak Physics and the early
Universe}, 23-25 March 1994, Sintra (Portugal).}}

\author{{\bf J. R. Espinosa} \thanks{Supported by a grant of Comunidad de
Madrid, Spain.}\\
Instituto de Estructura de la Materia, CSIC\\
Serrano 123, 28006-Madrid, Spain}

\date{}
\maketitle
\vspace{1.5cm}
\def\baselinestretch{1.15}
\begin{abstract}

We study the possibility of relaxing the cosmological bound on the
Higgs mass coming from the requirement of non-erasure of the baryon
asymmetry by sphalerons. After reviewing the Standard Model case
we obtain this bound in two extensions of it:
1) The Standard Model with an additional gauge singlet, and
2) The Minimal Supersymmetric Standard Model.
Taking fully into account all experimental constraints and thermal
screening effects we found that the situation can be slightly
improved with respect to the Standard Model but only in case 1) a non
negligible region in parameter space exists where the baryon washout
is avoided and the experimental bounds evaded.
\end{abstract}

\vskip-15.5cm
\rightline{{\bf hep-ph/9405319}}
\rightline{{\bf IEM--FT--85/94}}
\rightline{{\bf May 1994}}

\end{titlepage}


\section*{Introduction}

The attractive possibility that the observed baryon asymmetry of the
universe was created at the electroweak phase transition has
deserved a lot of of attention in the last years \cite{RE}.
It was realized \cite{KURUSH} that
all the conditions formulated by Sakharov \cite{SA} (for generating
the baryon asymmetry) can be met
in the Standard Model (SM). In particular, the sphaleron transitions were
recognized as unsupressed at high temperatures and so playing a
central role in the mechanisms generating the
amount of $\Delta B$ we finally observe today.

These sphaleron processes, if unsupressed after the electroweak phase
transition, will drive the previously created baryon asymmetry to
zero, and so its rate $\Gamma$ at that time \cite{KURUSH,ARLE}
($\Gamma \sim exp(-E_{sph}/T)$) should be less than the Hubble
expansion rate $H$ to go out of equilibrium.
Due to the exponential sensitivity of $\Gamma$ to the ratio
$E_{sph}(T)/T$, imposing $\Gamma < H$ gives the model independent
condition \cite{SH}:
\be
\label{central}
\frac{E_{sph}(T_c)}{T_c} \geq 45.
\ee

After relating the sphaleron energy \cite{KLMA} $E_{sph}$ to the vacuum
expectation value of the Higgs field at the critical
temperature $T_c$ ($E_{sph}(T_c)\sim \langle\phi(T_c)\rangle/g$) the central
relation Eq. (\ref{central}) translates into
\be
\label{strong}
\frac{\langle\phi (T_c)\rangle}{T_c} \simgt 1,
\ee
which tells that the transition should be strongly
enough first order so as to avoid the erasure of the baryon asymmetry.

When the dependence of $\langle\phi (T_c)\rangle$ on the parameters
of the Higgs potential is taken into account,
condition (2) gives an upper bound on the mass of the
Higgs boson (in the Standard Model where this calculation was
originally performed and also in more general cases).
The study of
this bound is the central aim of the work here resumed.

All along this paper we will say that a phase transition is strongly
first order if condition (\ref{strong}) is satisfied and weakly first
order (or even second order) if it is not.

\section*{The Standard Model}

The effective potential at non zero temperature is the central
tool in studying the electroweak phase transition. In one-loop
approximation the temperature-dependent
potential describes a first order phase transition between the
symmetric phase (at high T) and the $SU(2)\times U(1)$ breaking one.
Using a high temperature expansion the form of the potential is:
\be
V(\phi,T)=D(T^2-T_0^2)\phi^2-E T\phi^3+\frac{1}{4}\lambda(T)\phi^4,
\ee
where $D, T_0, E$ and $\lambda(T)$ are easily calculable.

Approximating the critical temperature $T_c$ as the temperature $T_D$
at which two degenerate minima
coexist ($T_c$ will be somewhere in the narrow range
between $T_D$ and the temperature $T_0$ at which the
minimum at $\phi=0$ is destabilized) one gets
\be
\label{MAIN}
\frac{\langle\phi (T_c)\rangle}{T_c} = \frac{2  E}{\lambda(T_c)}.
\ee
We see from this formula that the strenght of the transition is
governed by $E$ (which gives the cubic term in the potential).
Now, $E$ comes from bosonic loops only (from $n=0$ Matsubara
frecuencies). Each bosonic degree of freedom, with field dependent
mass $M_i(\phi)$ contributes to the
potential a term
\be
\Delta_i V = -\frac{T}{12 \pi}\left[ M_i^2(\phi)\right]^{3/2}.
\ee
In the SM, the main contribution is the one coming from gauge boson
loops:
\be
\label{GB}
\Delta V = -\frac{T}{12 \pi}\left[ 6 \left(
\frac{1}{4}g^2\phi^2\right)^{3/2} +  3 \left(
\frac{1}{4}(g^2+g'^2)\phi^2\right)^{3/2}\right].
\ee
Using the value of $E$ obtained from Eq. (\ref{GB}) and relating
$\lambda$ to the Higgs mass one gets the bound \cite{SH,DIHUSI}
\be
M_h\simlt  (45 - 50)\ GeV,
\ee
region which is already ruled out by LEP.

This bound is obtained within the one-loop approximation to the
effective potential and as is well known this will only be reliable
for $\phi>T$ (the expansion parameter is $g^2 T^2 / M^2(\phi)$). For
small $\phi$ the perturbative expansion is plagued with infrared
divergences. To cure this, one has to reorganize the loop expansion
resumming the most important infrared divergent
diagrams \cite{CASH} (the so
called daisy diagrams). The effect of this daisy resummation is to
shift the mass of the $n=0$ modes contributing to the cubic term by a
thermal mass (Debye screening effect).
In the SM this thermal mass is of order $g^2 T^2$ to leading order
for the longitudinal gauge bosons, and zero (to this order) for
the transverse ones.
The effect of this thermal mass is to shield the cubic term of the
longitudinal modes while leaving the transverse modes
unaffected and so to effectively reduce $E$ by a factor $2/3$.
The improved bound on the Higgs mass is then even lower \cite{DHLLL}:
\be
M_h\simlt  \sqrt{2/3} (45 - 50)\ GeV\sim (35 - 40)\ GeV.
\ee

There exist further refinements of this calculation in the
literature. If the screening of the transverse modes at subleading
order $(g^4 T^2)$ is taken into account \cite{MM} the strenght
of the transition is further reduced.
On the other hand, some non-perturbative effects may give a lower
value for the critical temperature \cite{KRS}
translating into a larger ratio
$\langle\phi (T_c)\rangle/T_c$. Also there are some two-loop
(resummed) calculations \cite{TWOL}
that give larger $\langle\phi (T_c)\rangle$.

Even with these uncertainties the SM has another problems to be able
to generate the observed baryon asymmetry. In fact the $CP$ violation
effects are by far too small to give the correct amount of $\Delta B$
and some recent atempts \cite{FS} to solve this problem have been
shown \cite{GA} to be flawn. So it seems necessary to
move to some extensions of the SM to explain the generation of the
baryon asymmetry at the electroweak phase transition.

{}From what we have learned by studying the Standard Model case we
can work out an estrategy \cite{ANHA} for the extensions of it
to give a strong first order phase transition.
A look at Eq. (\ref{MAIN}) shows that we need a large cubic term
in our potential.
So, adding more bosonic degrees of freedom may help for only bosons contribute
to this cubic term.
The screened mass of such bosons will have the
general form
\be
\label{GENMASS}
\overline{M}^2(\phi) = M^2 + \lambda \phi^2 + \Pi (T),
\ee
where $M$ is an invariant mass, $\lambda$ is a generic coupling of
the new bosons to the Higgs (here $\phi$ stands generically for the
fields driving the transition) and $\Pi$ is the thermal mass.
For this to give a non-negligible contribution to the cubic term,
small values of $M$ and $\Pi$ and large values of $\lambda$ are
required. Usually, there will exist an experimental lower bound for
$M$ so that one cannot take it to be arbitrarily small.
Concerning the coupling $\lambda$ the requirement of perturbativity
up to some large scale will set an upper bound on it. In a given
model one has to take all this conditions into account when looking
for a region in the parameter space where the phase transition is
strong enough.

Another way of increasing the strenght of the transition has
been considered in the
literature \cite{ANHA,BENSON,ZHANG}, namely
the possibility of decreasing the value of the coupling in the
denominator of Eq. (\ref{MAIN}) due to one loop corrections (even at zero
temperature). But usually for this
mechanism to give a sizeable effect large couplings
and/or many bosonic degrees of freedom are needed.

Before moving to the discussion of extended models some words about
the validity of the perturbative calculations are needed. Roughly
speaking, one can trust the (resummed) perturbative calculations in
the SM if $m_h^2\simlt m_W^2$ (or equivalently $\lambda / g^2 \simlt
1$). When trying to improve the "sphaleron bound"
$m_h\simlt m_{h,crit}$, one can face cases where $m_{h,crit}\sim m_W$
rising some doubts about the reliability of the calculation of
$m_{h,crit}$. In the cases we are going to consider the relevant
parameter will no longer be $\lambda/g^2$. In fact $g^2$ will be
substituted by $\zeta^2$ in the SM + singlet case and by $h_t^2$
in the MSSM . As we will be interested in large values of these
couplings, for a given Higgs mass the expansion parameter will be
smaller than in the SM, and the perturbative calculation more reliable.

\section*{The Standard Model with a gauge Singlet}

This is the simplest extension of the Standard Model that can in principle
improve the situation concerning the electroweak baryogenesis
problem, as already noted in Ref. \cite{ANHA}.

The lagrangian of the model is defined as:
\begin{equation}
\label{lag}
{\cal L}={\cal L}_{SM}+\partial^{\mu}S^*\partial_{\mu}S-
M^2S^*S-\lambda_S(S^*S)^2-2\zeta^2 S^*SH^*H,
\end{equation}
where $H$ is the SM doublet with $\langle H
\rangle=\phi/\sqrt{2}$, $\phi$ is the classical field, and $M^2,
\lambda_S, \zeta^2 \ge 0$, to guarantee that $\langle S\rangle=0$ at all
temperatures.

The temperature dependent effective potential in the one loop
approximation was studied in Ref. \cite{ANHA}. The daisy improvement
was performed in Refs. \cite{EQ,BENSON}. The potential is a
function of the classical field $\phi$ as in the SM, but with
a new contribution coming from the $S$ bosonic
loops of the form (imposing renormalization conditions preserving the tree
level value of the $vev$ $v$)
\be
\Delta V_S=g_S
\left\{ \frac{m_S^2(\phi)T^2}{24}-\frac{\overline{
M}_S^3(\phi)T}{12\pi} -\frac{m_S^4(\phi)}{64\pi^2}
\left[\log\frac{m_S^2(v)}{c_BT^2}-2\frac{m_S^2(v)}{m_S^2(\phi)}
\right] \right\},
\ee
where $g_S=2$ is the number of degrees of freedom of the (complex) singlet.
$m_S(\phi)$ is the field dependent mass of $S$
and $\overline{M}_S(\phi)$ its Debye screened mass. They are given by
\bear{ccl}
m_S^2(\phi)&=&M^2+\zeta^2 \phi^2,\vspace{.5cm}\\
\overline{M}_S^2&=&m_S^2(\phi)+\Pi_S(T).
\eear
Here $\Pi_S$ is, to leading order
\be
\Pi_S(T)=\frac{\lambda_S+\zeta^2}{3}T^2.
\ee
The thermal polarizations for the rest of SM particles can receive
extra contributions coming from $S$ loops (see Refs. \cite{EQ,BENSON}).

To understand the numerical results presented at the end of this
section it is convenient to study analytically the effective potential
assuming as will be the case that the bosonic contribution is
dominated by the singlet (and this is the reason for the relevant
expansion parameter to be $\lambda/\zeta^2$ rather than
$\lambda/g^2$). The high temperature expansion of the
effective potential takes then the form
\be
V(\phi)=A(T)\phi^2+B(T)\phi^4+C(T)\left(\phi^2+K^2(T)\right)^{3/2},
\ee
where
\begin{equation}
\label{C}
C(T)=-\frac{\zeta^3T}{6\pi},
\end{equation}
\begin{equation}
\label{K2}
K^2(T)=\frac{(\zeta^2+\lambda_S)T^2+3M^2}{3\zeta^2}.
\end{equation}
{}From these expressions and the discussion at the end of the previous
section is clear that the best case for the phase transition to be
stongly first order is to have low values of $M$ and $\lambda_S$ and
large values of $\zeta^2$. This is what was found numerically in
Ref. \cite{EQ}.

For large values of $M$ the transition is
weaker and eventually, when $M\gg T$,
the singlet decouples and the SM
result is recovered. The dependence
on $\zeta^2$ is the most interesting one. Imposing the condition that
all the couplings in the theory remain perturbative up to some high
scale $\Lambda$ an upper bound for $\zeta^2$ at the electroweak scale
is derived (as a function of course of the scale $\Lambda$ and the
unknown value of the top yukawa coupling $h_t$). This renormalization
group study was also performed in Ref. \cite{EQ}.

Translating the requirement of a strong phase transition
($E_{sph}(T_c)/T_c>45$ or $\langle\phi (T_c)\rangle\simgt 1.3$ with our
definition of $T_c$) into a bound
on the Higgs mass $M_H^{crit}$, one finds that in this model
one can evade
the stringent bound of the SM and go to larger values of
$M_H^{crit}$. The numerical value of this critical mass depends on
how large the coupling $\zeta^2$ is, or equivalently, on
the mass of the top and the scale of new physics $\Lambda$. Assuming
that this scale is not lower than $10^6\ GeV$  the value of
$M_H^{crit}$ can be of order $80\ GeV$ (see Fig. 1) which is higher than the
experimental bound. Larger values of this critical mass can only be
obtained at the price of introducing new physics at lower scales.

\section*{The\, Minimal\, Supersymmetric\, Standard Model}

The Minimal Supersymmetric Standard Model (MSSM) is the physically
most motivated and phenomenologically most acceptable among the
extensions of the SM.
Concerning the generation of the baryon asymmetry the MSSM allows
for extra CP-violating phases besides the Kobayashi-Maskawa one,
which could help in generating the observed baryon
asymmetry \cite{CONE}.
On the other hand the MSSM has a lot of new degrees of freedom and
the nature of the phase
transition could be significantly modified with respect to the SM.

The main tool for our study is the one-loop, daisy-improved (for
previous studies in the one-loop approximation see
Refs. \cite{DIHUSI,MSSM1})
finite-temperature
effective potential of the MSSM, $V_{\rm{eff}}(\phi,T)$. Now the
potential is actually a function of two fields: $\phi_1 \equiv {\rm
Re}
\, H_1^0$ and $\phi_2 \equiv {\rm Re} \,  H_2^0$, where $H_1^0$
and $H_2^0$ are the neutral
components of the Higgs doublets $H_1$ and $H_2$, thus $\phi$ will
stand
for $(\phi_1,\phi_2)$. Working in the 't~Hooft-Landau gauge and in
the
$\ov{DR}$-scheme, we can write
\be
\label{total}
V_{\rm{eff}}(\phi,T) = V_0(\phi)
+ V_1(\phi,0) + \Delta V_1(\phi,T)
+\Delta V_{\rm{daisy}}(\phi,T) \, ,
\ee
where
\bea
\label{v0}
V_0(\phi) & = &  m_1^2 \phi_1^2 + m_2^2  \phi_2^2 + 2 m_3^2 \phi_1
\phi_2
+ {g^2+g'\,^2 \over 8} (\phi_1^2 -\phi_2^2)^2 \, ,
\vspace*{.5cm}\\
\label{deltav}
V_1(\phi,0) & = & \sum_i {n_i \over
64 \pi^2} m_i^4 (\phi) \left[ \log {m_i^2 (\phi) \over Q^2} - {3
\over 2} \right]  \, ,
\vspace*{.5cm}\\
\label{deltavt}
\Delta V_1(\phi,T) & = & {T^4 \over 2 \pi^2} \left\{ \sum_i
n_i \, J_i \left[ { m^2_i (\phi) \over T^2 } \right]  \right\} \, ,
\vspace*{.5cm}\\
\label{dvdaisy}
\Delta V_{\rm{daisy}}(\phi,T) & = & - {T \over 12 \pi} \sum_i n_i
\left[ \ov{m}_i^3 (\phi, T ) - m_i^3 (\phi) \right] \, .
\eea
The first term,
Eq.~(\ref{v0}),
is the tree-level potential. The second term, Eq.~(\ref{deltav}), is
the one-loop contribution at $T=0$: $Q$ is the renormalization scale,
where we
choose for
definiteness $Q^2 = m_Z^2$, $m_i^2 (\phi)$ is the field-dependent
mass of
the $i^{th}$ particle, and $n_i$ is the corresponding number of
degrees of
freedom, taken negative for fermions. Since $V_1(\phi,0)$ is
dominated by
top ($t$) and stop ($\st_1,\st_2$) contributions, only these will be
included
in the following. The third term, Eq.~(\ref{deltavt}), is the
additional
one-loop
contribution due to temperature effects. Here $J_i=J_+ (J_-)$ if the
$i^{th}$
particle is a boson (fermion), and
\be
\label{ypsilon}
J_{\pm} (y^2) \equiv \int_0^{\infty} dx \, x^2 \,
\log \left( 1 \mp e^{- \sqrt{x^2 + y^2}} \right) \, .
\ee
Since the relevant contributions to $\Delta V_1(\phi,T)$ are due to
top ($t$),
stops ($\st_1,\st_2$) and gauge bosons ($W,Z$), only these will be
considered
in the following. Finally, the last term, Eq.~(\ref{dvdaisy}), is the
correction
coming from daisy diagrams. The sum
runs over bosons only. As usual, the masses $\ov{m}_i^2 (\phi,T)$ are obtained
from the ${m}_i^2 (\phi)$ by adding the leading $T$-dependent
self-energy
contributions, which are proportional to $T^2$.

The relevant degrees of freedom for our calculation are:
\be
\label{multi}
n_t = - 12 \, ,  \;\;
n_{\st_1} = n_{\st_2} = 6 \,  , \;\;
n_W=6 \,  , \;\; n_Z=3 \,  , \;\;
n_{W_L}=2 \, ,  \;\; n_{Z_L}=n_{\gamma_L}=1 \, .
\ee
The field-dependent top mass is
\be
\label{tmass}
m_t^2(\phi)=h_t^2 \phi_2^2 \, .
\ee
The entries of the field-dependent stop mass matrix are
\bea
\label{tlmass}
m_{\tilde{t}_L}^2 (\phi) & = & m_{Q_3}^2 + m_t^2 (\phi) +
D_{\tilde{t}_L}^2 (\phi)  \, ,
\vspace{.5cm}\\
\label{trmass}
m_{\tilde{t}_R}^2 (\phi) & = & m_{U_3}^2 + m_t^2 (\phi) +
D_{\tilde{t}_R}^2 (\phi)  \, ,
\vspace{.5cm}\\
\label{mixmass}
m_X^2 (\phi) & = & h_t (A_t \phi_2 + \mu \phi_1) \, ,
\eea
where $m_{Q_3}$, $m_{U_3}$ and $A_t$ are soft supersymmetry-breaking
mass
parameters, $\mu$ is a superpotential Higgs mass term, and
\bea
\label{dterms}
D_{\tilde{t}_L}^2(\phi)& = & \left( {1 \over 2}-
{2 \over 3}\sin^2 \theta_W \right)
{g^2 + g'\,^2 \over 2}(\phi_1^2-\phi_2^2),\\
D_{\tilde{t}_R}^2(\phi) &=&\left( {2 \over 3}\sin^2 \theta_W \right)
{g^2 + g'\,^2 \over 2}(\phi_1^2-\phi_2^2)
\eea
are the $D$-term contributions. The field-dependent stop masses are
then
\be
\label{mstop}
m_{\tilde{t}_{1,2}}^2 (\phi) = {m^2_{\tilde{t}_L} (\phi) +
m^2_{\tilde{t}_R} (\phi) \over 2} \pm \sqrt{ \left[
{m^2_{\tilde{t}_L} (\phi)- m^2_{\tilde{t}_R} (\phi) \over 2}
\right]^2 + \left[ m_X^2(\phi) \right]^2  } \, .
\ee
The corresponding effective $T$-dependent masses,
$\ov{m}^2_{\tilde{t}_{1,2}}
(\phi,T)$, are given by expressions identical to (\ref{mstop}), apart
from the
replacement
\be
\label{repl}
m^2_{\tilde{t}_{L,R}} (\phi) \, \rightarrow \,
\ov{m}^2_{\tilde{t}_{L,R}}(\phi,T) \equiv
m^2_{\tilde{t}_{L,R}} (\phi)+  \Pi_{\tilde{t}_{L,R}}(T)  \,  .
\ee
The $\Pi_{\tilde{t}_{L,R}}(T)$ are the leading parts of the
$T$-dependent
self-energies of $\tilde{t}_{L,R}$\,,
\bea
\label{pistl}
\Pi_{\tilde{t}_L}(T)& = &
{4 \over 9}g_s^2 T^2 +
{1 \over 4}g^2 T^2 +
{1 \over 108}g'\,^2  T^2 +
{1 \over 6}h_t^2 T^2 \, ,
\vspace{.5cm}\\
\label{pistr}
\Pi_{\tilde{t}_R}(T) & =  &
{4 \over 9}g_s^2 T^2 +
{4 \over 27} g'\,^2 T^2 +
{1 \over 3}h_t^2 T^2 \, ,
\eea
where $g_s$ is the strong gauge coupling constant. Only loops of
gauge
bosons, Higgs bosons and third generation squarks have been included,
implicitly assuming that all remaining supersymmetric particles are
heavy and decouple. If some of these are also light, the plasma
masses for the stops  will be even larger, further suppressing the
effects of the associated cubic terms, and therefore weakening the
first-order nature of the phase transition.

We expect the stops to play an important role in making the
transition stronger as they have a large number of degrees of freedom
and couple to the Higgses with strength $h_t$. On the other hand as they are
coloured scalars their thermal mass is of order $g_s^2 T^2$ as shown
above and also they are not protected by any symmetry to have an
invariant mass. The final importance they have in the strenght of the
transition will result from the interplay between these opposite
properties.

Finally, the
field-dependent
gauge boson masses are
\be
\label{gauge}
m_W^2 (\phi) = {g^2 \over 2} (\phi_1^2 + \phi_2^2)  \, ,
\;\;\;\;\;\;
m_Z^2 (\phi) = {g^2 + g'\,^2 \over 2} (\phi_1^2 + \phi_2^2)  \, ,
\ee
and the effective $T$-dependent masses of the longitudinal gauge
bosons
are
\bea
\label{wmass}
\phantom{b}&\phantom{b}&
\ov{m}_{W_L}^2(\phi,T) \ \  =  m_W^2 (\phi)+ \Pi_{W_L}(T)  \, ,
\\
\label{zphlmass}
\phantom{b}&\phantom{b}&
\ov{m}^2_{Z_L,\gamma_L}(\phi,T)  =
\frac{1}{2} \left[ m_Z^2 (\phi) + \Pi_{W_L}(T) + \Pi_{B_L}(T) \right]
\nonumber\\
& \pm &
\sqrt{ \frac{1}{4} \left[ {g^2 - g'\,^2 \over 2} (\phi_1^2 +
\phi_2^2)
+ \Pi_{W_L}(T) - \Pi_{B_L}(T) \right]^2
+ \left[ {gg' \over 2}(\phi_1^2 + \phi_2^2) \right]^2 } \, .
\eea
In eqs.~(\ref{wmass}) and (\ref{zphlmass}), $\Pi_{W_L}(T)$ and
$\Pi_{B_L}(T)$
are the leading parts of the $T$-dependent self-energies of $W_L$ and
$B_L$,
given by
\be
\label{piwb}
\Pi_{W_L}(T)  =  {5 \over 2}g^2 T^2 \, ,
\;\;\;\;\;
\Pi_{B_L}(T) =   {47 \over 18}g'\,^2 T^2 \, ,
\ee
where only loops of Higgs bosons, gauge bosons, Standard Model
fermions
and third-generation squarks have been included.

We need to analyse the effective potential (\ref{total}) as a
function of
$\phi$ and $T$. Before doing this, however, we trade the parameters
$m_1^2,
m_2^2,m_3^2$ appearing in the tree-level potential (\ref{v0}) for
more
convenient parameters. To this purpose, we first minimize the
zero-temperature
effective potential, i.e. we impose the vanishing of the first
derivatives of
$V_0(\phi) + V_1(\phi,0)$ at $(\phi_1,\phi_2)=(v_1,v_2)$, where
$(v_1,v_2)$
are the one-loop vacuum expectation values at $T=0$. This allows us
to eliminate
$m_1^2$ and $m_2^2$ in favour of $m_Z^2$ and $\tan\beta\equiv
v_2/v_1$ in the standard way.
Moreover, $m_3^2$ can be traded for the one-loop-corrected mass
$m_A^2$
of the CP-odd neutral Higgs boson \cite{erz2}.
Therefore the whole effective potential (\ref{total}) is completely
determined, in our approximation, by the parameters $(m_A,\tb)$ of
the Higgs sector, and by the parameters ($m_t$, $m_{Q_3}$, $m_{U_3}$,
$\mu$, $A_t$) of the top/stop sector. The same set of parameters also
determines the one-loop-corrected masses and couplings of the MSSM
Higgs bosons.

The next steps are the computation of the critical temperature and
of the location of the minimum of the effective potential at the
critical temperature.
We define here $T_0$ as the temperature at which the determinant of
the
second derivatives of $V_{\rm eff}(\phi,T)$ at $\phi=0$ vanishes:
\be
\label{det}
\det\left[
{{\partial^2 V_{\rm eff}(\phi,T_0)}\over{\partial \phi_i \partial
\phi_j}}
\right]_{\phi_{1,2}=0} = 0  \, .
\ee
It is straightforward to compute the derivatives in eq.~(\ref{det})
from the previous formulae; the explicit expressions are given in
Ref. \cite{beqz}.

Once eq.~(\ref{det}) is solved (numerically) and $T_0$ is found, one
can
minimize (numerically) the potential $V_{\rm eff}(\phi,T_0)$ and find
the
minimum $[v_1(T_0),v_2(T_0)]$. The quantity of interest is indeed, as
will
be discussed later, the ratio $v(T_0)/T_0$, where $v(T_0)\equiv
\sqrt{v_1^2
(T_0)+v_2^2(T_0)}$.

Before presenting the numerical results we discuss the experimental
constraints on the parameters of the top/stop sector and of the Higgs
sector. We treat $m_{Q_3}$, $m_{U_3}$ and the other soft mass terms as
independent parameters, even if they can be related in specific
SUGRA models. We want to be as general as possible and in this way
the region of parameters we are able to exclude will be excluded in
any particular model.

Direct and indirect searches
at LEP imply \cite{coignet} that $m_{\tilde{b}_L} \simgt 45 \gev$,
which
in turn translates into a  bound in the $(m_{Q_3},\tb)$ plane.
Electroweak
precision measurements \cite{altarelli} put stringent constraints on
a light
stop-sbottom sector: in first approximation, and taking into account
possible
effects \cite{abc} of other light particles of the MSSM, we
conservatively
summarize the constraints by $\Delta \rho (t,b) + \Delta \rho
(\tilde{t},
\tilde{b}) < 0.01$.

We finally need to consider the constraints coming from LEP searches
for
supersymmetric Higgs bosons \cite{coignet}.
Experimentalists put limits on the processes $Z \rightarrow h Z^*$
and
$Z \rightarrow h A$, where $h$ is the lighter neutral CP-even boson.
We
need to translate these limits into exclusion contours in the $\matb$
plane,
for given values of the top/stop parameters. In order to do this, we
identify
the value of $BR(Z \rightarrow h Z^*)$, which corresponds to the
limit
$m_{\phi} > 63.5 \gev$ on the SM Higgs, and the value of $BR(Z
\rightarrow
h A)$, which best fits the published limits for the representative
parameter
choice $m_t = 140 \gev$, $m_{Q_3} = m_{U_3} \equiv \tilde{m} = 1
\tev$,
$A_t = \mu = 0$. We then compare those values of $BR(Z \rightarrow h
Z^*)$
and $BR(Z \rightarrow h A)$ with the theoretical predictions of the
MSSM,
for any desired parameter choice and after including the radiative
corrections
associated to top/stop loops \cite{pioneer,erz2}. Of course,
this procedure is not entirely  correct, since it ignores the
variations of
the efficiencies  with the Higgs masses and branching ratios, as well
as
the possible presence of candidate events at some mass values, but it
is adequate for our purposes.

We now present our numerical results, based on the effective
potential of eq.~(\ref{total}), concerning the strength of the
electroweak phase transition and the condition for preserving the
baryon asymmetry.

Particularizing to the MSSM the
studies
of sphalerons in general two-Higgs models \cite{two}, we obtain
that
\be
E_{sph}^{MSSM} (T) \le E_{sph}^{SM} (T) \, ,
\ee
where, in our conventions,
\be
\label{esph}
{E_{sph}^{SM} (T) \over T} = {4 \sqrt{2} \pi \over g} B
\left\{ {\lambda_{\rm eff} (T) \over 4 g^2 } \right\} {v(T)
\over T} \, ,
\ee
and $B$ is a smoothly varying function whose values can be found in
Ref. \cite{KLMA}. For example, $B(10^{-2})=1.67$, $B(10^{-1})=1.83$, $B(1)
=2.10$. It can also be shown that
\be
{v(T_D) \over T_D} < {v(T_C) \over T_C} <
{v(T_0) \over T_0} \, ,
\ee
where $T_C$ is the actual temperature at which the phase transition
occurs, satisfying the inequalities
\be
T_0 < T_C < T_D \, ,
\ee
if $T_0$ is defined by (\ref{det}) and $T_D$ is the temperature at
which
there are two degenerate minima.

Finally, the corrections in $E_{sph}^{SM}$ due to $g'\neq 0$
have been estimated and shown to be
small \cite{mixing}. Therefore, a conservative bound to be imposed
is
\be
\label{erre}
R \equiv {v(T_0) \over T_0} {4 \sqrt{2} \pi  B
\left\{ {\lambda_{\rm eff} (T_0) \over 4 g^2 } \right\} \over
45 g} > 1 \, .
\ee

The last point to be discussed is the determination of the value of
$\lambda_{\rm eff} (T_0)$ to be plugged into
eq.~(\ref{erre}). The $B$-function we use is taken from
Ref.~\cite{KLMA}, where the sphaleron energy was computed using the
zero-temperature `Mexican-hat' potential, $V = \frac{\lambda}{4}
(\phi^2-v^2)^2$. The sphaleron energy at finite temperature was
computed in Ref.~\cite{belgians}, where it was proven that it scales
like $v(T)$, i.e.
\begin{equation}
\label{aaa}
E^{SM}_{sph}(T)=E^{SM}_{sph}(0) \; \frac{v(T)}{v} \, ,
\end{equation}
with great accuracy. Therefore, to determine the value of
$\lambda_{\rm eff} (T_0)$ we have fitted $V_{\rm
eff}(\phi,T_0)$, in the direction of the minimum,
to the one-dimensional approximate
potential,
\begin{equation}
\label{bbb}
V_{\rm eff}(\phi,T_0) \simeq \frac{1}{4} \lambda_{\rm
eff}(T_0)
[\phi^2-v^2(T_0)]^2 \,
,
\end{equation}
The value
of $\lambda_{\rm eff}$ obtained from (\ref{bbb}),
\begin{equation}
\label{ccc}
\lambda_{\rm eff}(T_0)=4 \;
\frac{V_{\rm eff}(0,T_0)-V_{\rm eff}[v(T_0),T_0]} {v^4(T_0)} \, ,
\end{equation}
where all quantities on the right-hand side are calculated
numerically
from the potential of eq.~(\ref{total}), is then plugged into
eq.~(\ref{erre}) to obtain our bounds. Of course, the quality of the
fit is good only for values of $\phi \simlt v(T_0)$ but this is
precisely the region of interest to determine the sphaleron energy.

Our numerical results are summarized in fig.~2, in the $\matb$ plane
and for two representative values of the top quark mass: $m_t = 130
\gev$ (fig.~2a) and $m_t = 170 \gev$ (fig.~2b). In each case, the
values of the remaining free parameters have been chosen in order
to maximize the strength of the phase transition, given the
experimental constraints on the top-stop sector. Notice that
arbitrarily small values
of $m_{U_3}$ cannot be excluded on general grounds, even if they are
disfavoured by model calculations. Also, we have explicitly checked
that,
as in Ref.~\cite{eqz3}, mixing effects in the stop mass matrix
always
worsen the case. In fig.~2, solid lines correspond to contours of
constant $R$: one can see that the requirement
of large values of $R$ favours small
$\tb$ and $m_A \gg m_Z$. The thick solid line corresponds to the
limits coming from Higgs searches at LEP: for our parameter choices,
the
allowed regions correspond to large $\tb$ and/or $m_A \gg m_Z$. For
reference, contours of constant $m_h$ (in GeV) have also been plotted
as dashed lines. One can see that, even for third-generation
squarks as
light as allowed by all phenomenological constraints,
only a very small globally allowed region can exist
in the $\matb$ plane, and that the most favourable
situation
is the one already discussed in Ref.~\cite{eqz3}. More precisely,
the region that is still marginally allowed corresponds to $m_A \gg
m_Z$, $\tb \sim 2$, stop and sbottom sectors as light as otherwise
allowed, a heavy top, and a light Higgs boson with SM-like properties
and
mass $m_h \sim 65 \gev$, just above the present experimental limit.
A less conservative interpretation of the limits from precision
measurements, the inclusion of some theoretically motivated
constraints on the model parameters, or a few GeV improvement in the
SM Higgs mass limit, would each be enough to fully exclude
electroweak baryogenesis in the MSSM.

\section*{Acknowledgments}

The results presented here are based on joint work done in collaboration with
A.~Brignole, M.~Quir\'os and F.~Zwirner. I must say it is always a pleasure
to work with them.

\section*{Figure captions}
\begin{itemize}
\item[Fig.1:]
Plot of $\langle\phi(T_c)\rangle/T_c$ versus $m_h$ for $\lambda_S=0$, $M=0$,
$m_t=90,\ 150$ and $175\ GeV$ and $\zeta$ corresponding to
$\Lambda= 10^6\ GeV$.
\item[Fig.2:]
Contours of $R$ in the $\matb$ plane, for the parameter choices:
a) $m_t$ = 130 $\gev$, $m_{Q_3}$ = 50 $\gev$, $m_{U_3}$ = 0
($m_{\tilde{t}}$ $\sim$ 130 $\gev$, $m_{\tilde{b}_L}$ $\sim$ 50 $\gev$),
$A_t$ = $\mu$ = 0;
b) $m_t$ = 170 $\gev$, $m_{Q_3}$ = 280 $\gev$, $m_{U_3}$ = 0
($m_{\tilde{t}_L}$ $\sim$ 330 $\gev$, $m_{\tilde{t}_R}$ $\sim$ 170 $\gev$,
$m_{\tilde{b}_L}$ $\sim$ 280 $\gev$),
$A_t$ = $\mu$ = 0.
The region excluded by Higgs searches at LEP is delimited by the
thick solid line. For reference, contours of constant $m_h$ (in GeV)
are also represented as dashed lines.
\end{itemize}


\begin{thebibliography}{99}
%
%
\bibitem{RE}
For reviews and references see, e.g.:
\\
A.D.~Dolgov, {\em Phys. Rep.} {\bf 222} (1992) 309;
A.G.~Cohen, D.B.~Kaplan and A.E.~Nelson, {\em Ann. Rev. Nucl. Part. Sci.}
{\bf 43} (1993) 27;
M.~Dine, preprint SCIPP 94/03.
%
%
\bibitem{KURUSH}
V.A.~Kuzmin, V.A.~Rubakov and M.E.~Shaposhnikov, {\em Phys.
Lett.} {\bf B155} (1985) 36.
%
%
\bibitem{SA}
A.D.~Sakharov, {\em JETP Lett.} {\bf 5} (1967) 24.
%
%
\bibitem{ARLE}
P.~Arnold and L.~McLerran, {\em Phys. Rev.} {\bf D36} (1987)
581; {\em Phys. Rev.} {\bf D37} (1988) 1020.
%
%
\bibitem{SH}
M.E.~Shaposhnikov, {\em JETP Lett.} {\bf 44} (1986) 465; {\em Nucl.
Phys.} {\bf B287} (1987)
757 and {\bf B299} (1988) 797; A.I.~Bochkarev and M.E.~Shaposhnikov,
{\em Mod. Phys. Lett.} {\bf A2} (1987) 417.
%
%
\bibitem{KLMA}
F.R.~Klinkhamer and N.S.~Manton, {\em Phys. Rev.} {\bf D30} (1984) 2212.
%
%
\bibitem{DIHUSI}
M.~Dine, P.~Huet and R.~Singleton Jr., {\em Nucl. Phys.} {\bf
B375} (1992) 625.
%
%
\bibitem{CASH}
M.E.~Carrington, {\em Phys. Rev.} {\bf D45} (1992) 2933;
M.E.~Shaposhnikov, {\em Phys. Lett.} {\bf B277}
(1992) 324 and (E) {\bf B282} (1992) 483;
C.G.~Boyd, D.E.~Brahm and S.D.H.~Hsu, {\em Phys. Rev.} {\bf D48} (1993) 4952.
%
%
\bibitem{DHLLL}
M.~Dine, R.G.~Leigh, P.~Huet, A.~Linde and D.~Linde, {\em Phys. Lett.}
{\bf B283}
(1992) 319 and {\em Phys. Rev.} {\bf D46} (1992) 550.
%
%
\bibitem{MM}
J.R.~Espinosa, M.~Quir\'os and F.~Zwirner, {\em Phys. Lett.} {\bf B314} (1993)
206;
W.~Buchm\"{u}ller, Z.~Fodor, T.~Helbig and D.~Walliser, preprint DESY
93-021.
%
%
\bibitem{KRS}
K.~Kajantie, K.~Rummukainen and M.E.~Shaposhnikov, {\em Nucl. Phys.} {\bf B407}
(1993) 356;
M.E.~Shaposhnikov, {\em Phys. Lett.} {\bf B316} (1993) 112.
%
%
\bibitem{TWOL}
P.~Arnold and O.~Espinosa, {\em Phys. Rev.} {\bf D47} (1993) 3546;
J.~Bagnasco and M.~Dine, {\em Phys. Lett.} {\bf B303} (1993) 308.
%
%
\bibitem{FS}
G.~Farrar and M.E.~Shaposhnikov, {\em Phys. Rev. Lett.} {\bf 70} (1993) 2833 +
(E) {\bf 71} (1993) 210 and preprint CERN-TH.6734/93.
%
%
\bibitem{GA}
M.B.~Gavela, P.~Hern\'andez, J.~Orloff and O.~P\`ene, preprint
CERN-TH.7081/93; P.~Huet and E.~Sather, preprint
SLAC-PUB-6479.
%
%
\bibitem{ANHA}G.W.~Anderson and L.J.~Hall, {\em Phys. Rev.} {\bf
D45} (1992) 2685.
%
%
\bibitem{BENSON}K.E.C.~Benson, {\em Phys. Rev.} {\bf D48} (1993) 2456.
%
%
\bibitem{ZHANG}X.~Zhang, preprint UMDHEP 93-074.
%
%
\bibitem{EQ}J.R.~Espinosa and M.~Quir\'os, {\em Phys. Lett.} {\bf
B305} (1993) 98.
%
%
\bibitem{CONE}A.G.~Cohen and A.E.~Nelson, {\em Phys. Lett.} {\bf B297}
(1992) 111.
%
%
\bibitem{MSSM1}
G.F.~Giudice, {\em Phys. Rev.} {\bf D45} (1992) 3177;
S.~Myint, {\em Phys. Lett.} {\bf B287} (1992) 325.
%
%
\bibitem{erz2}
J.~Ellis, G.~Ridolfi and F.~Zwirner, {\em Phys. Lett.} {\bf B262} (1991) 477.
%
%
\bibitem{beqz}
A. Brignole, J.R.~Espinosa, M.~Quir\'os and F.~Zwirner,
{\em Phys. Lett.} {\bf B324} (1994) 181.
%
%
\bibitem{coignet}
G.~Coignet, Plenary talk at the XVI International Symposium on
Lepton-Photon Interactions, Cornell University, Ithaca, New York,
10--15 August 1993.
%
%
\bibitem{altarelli}
G.~Altarelli, Plenary talk given at the International
Europhysics Conference on High Energy Physics, Marseille, 22--28 July
1993, preprint CERN-TH.7045/93.
%
%
\bibitem{abc}
R.~Barbieri, M.~Frigeni and F.~Caravaglios, {\em Phys. Lett.} {\bf B279} (1992)
169;
G.~Altarelli, R.~Barbieri and F.~Caravaglios, {\em Nucl. Phys.}
{\bf B405} (1993) 3, and {\em Phys. Lett.} {\bf B314} (1993) 357;
J.~Ellis, G.L.~Fogli and E.~Lisi, {\em Phys. Lett.} {\bf B285} (1992) 238,
{\bf B286} (1992) 85 and {\em Nucl. Phys.} {\bf B393} (1993) 3.
%
%
\bibitem{pioneer}
Y.~Okada, M.~Yamaguchi and T.~Yanagida, {\em Prog. Theor. Phys. Lett.}
{\bf 85} (1991) 1 and {\em Phys. Lett.} {\bf B262} (1991) 54;
J.~Ellis, G.~Ridolfi and F.~Zwirner, {\em Phys. Lett.} {\bf B257} (1991) 83;
H.E.~Haber and R.~Hempfling, {\em Phys. Rev. Lett.} {\bf 66} (1991) 1815;
R.~Barbieri and M.~Frigeni, {\em Phys. Lett.} {\bf B258} (1991) 395.
%
%
\bibitem{two}
A.~Bochkarev, S.~Kuzmin and M.~Shaposhnikov, {\em Phys. Rev.} {\bf D43} (1991)
369
and {\em Phys. Lett.} {\bf B244} (1990) 275;
B.~Kastening, R.D.~Peccei and X.~Zhang, {\em Phys. Lett.} {\bf B266}
(1991) 413.
%
%
\bibitem{mixing}
J.~Kunz, B.~Kleihaus and Y.~Brihaye, {\em Phys. Rev.} {\bf D46} (1992) 3587.
%
%
\bibitem{belgians}
S.~Braibant, Y.~Brihaye and J.~Kunz, {\em Int. J. Mod. Phys.} {\bf
A8} (1993) 5563.
%
\bibitem{eqz3}
J.R.~Espinosa, M.~Quir\'os and F.~Zwirner, {\em Phys. Lett.} {\bf B307} (1993)
106.
%
\end{thebibliography}
\end{document}